\documentclass{aa}  

\usepackage{graphicx}
\usepackage{txfonts}
\usepackage[allcolors=blue]{hyperref}
\usepackage{subfig}
\usepackage{tabularx}

\begin{document}
        
    \title{High-precision broadband linear polarimetry of early-type binaries \\IV. The DH Cephei binary system in the open cluster of NGC~7380\thanks{The polarization data for DH~Cep are only available in electronic form at the CDS via anonymous ftp to \url{cdsarc.cds.unistra.fr} (\url{130.79.128.5}) or via \url{https://cdsarc.cds.unistra.fr/cgi-bin/qcat?J/A+A/}}}
        
    \titlerunning{DH~Cep}
        
    \author{Yasir Abdul~Qadir\inst{1}
    \and  Andrei V. Berdyugin\inst{1}
    \and Vilppu Piirola\inst{1}\and\\
    Takeshi Sakanoi\inst{2}
    \and Masato Kagitani\inst{2}
    }
        
    \institute{Department of Physics and Astronomy, FI-20014 University of Turku, Finland \\
    \email{yasir.abdulqadir@utu.fi}
    \and Graduate School of Sciences, Tohoku University, Aoba-ku, 980-8578 Sendai, Japan 
    }
        
    \date{Received XX / Accepted XX}
        
    \abstract
    {}
    {DH~Cephei is a well-known massive O+O-type binary system on the northern sky, situated at the center of young open cluster NGC~7380. Our high-precision multi-band polarimetry clearly reveals that variations of linear polarization in this system are synchronous with the phase of the orbital period. We used the observed variations of Stokes parameters $q$ and $u$ to derive the orbital inclination $i$, orientation $\Omega$, and the direction of rotation. Moreover, in order to obtain a rough estimation of the interstellar polarization in the vicinity of DH Cep, we observed polarization arising from the neighboring stars in the cluster.}
    {We used the Dipol-2 polarimeter in combination with the remotely controlled 60 cm Tohoku T60 telescope to obtain linear polarization measurements of DH~Cep in the $B$, $V$, and $R$ passbands at the accuracy level of $\sim$0.003\%. To obtain an estimation of interstellar polarization of DH~Cep, we observed more than a dozen field stars identified as members of NGC~7380 and in the close proximity to DH~Cep. A Lomb-Scargle period search was applied to the acquired polarization data to reveal the dominating frequency in  polarization variations. We  used a standard analytical method based on a two-harmonics Fourier fit to derive the inclination, orientation, and the direction of rotation of the binary orbit.}
    {The variations of Stokes parameters in all three $B$, $V$, and $R$ passbands clearly suggest an unambiguous periodic signal at 1.055 d with an amplitude of variations of $\sim$$0.2\%,$ which corresponds to half of the known orbital period of 2.11 d. This type of polarization variability is expected for a binary system with light-scattering material distributed symmetrically with respect to the orbital plane. In addition to the regular polarization variability, there is a nonperiodic component, which is strongest in the $B$ passband.  In the $V$ passband, we obtained our most reliable values for the orbital inclination $i = 46^{\circ}+11^{\circ}/-46^{\circ}$ and an orientation of the orbit on the sky of $\Omega = 105^{\circ} \pm 55^{\circ}$, with 1$\sigma$ confidence intervals. Using our best estimate of $i$ and the polametric amplitude in the $V$ passband, we estimated that the mass loss from the system is $\sim$$3.4 \times 10^{-7}M_{\rm \odot} \; \rm yr^{-1}$. The direction of the binary system rotation on the plane of the sky is clockwise. Our polarimetric observations of neighboring stars of DH~Cep in NGC~7380 reveal that the polarization of
the cluster stars is most likley due to aligned interstellar dust in the foreground.}
{}

        
    \maketitle
        
    \section{Introduction}\label{sec:intro}
    DH~Cephei (DH~Cep, or HD 215835) is a close binary system that consists of two O-type stars. It is the second-brightest member of the young open cluster NGC~7380 \citep{1969A&A.....1..356U} and is located in the Perseus arm of the Galaxy at
    a distance of $\sim$2.9~kpc \citep{2021A&A...649A...1G}. A number of photometric data analyses have been carried out over recent years, all of them suggesting that DH~Cep is an elliptical variable with no signatures of eclipses (\citet{1971A&A....13...30M}; \citet{1981IBVS.1965....1W}; \citet{1986IBVS.2932....1L}). 

    The first spectroscopic studies of this binary star system were carried out in 1949, where DH~Cep  was found to be a double-lined spectroscopic binary with a circular orbit and an orbital period of $\sim$2.11~d \citep{1949AJ.....54..135P}. \citet{1986IBVS.2932....1L} used their photometric observations to show that the photometric minima are separated by half an orbital period, reconfirming the orbital period of 2.11~d. The latest photometric study of DH~Cep by \citet{2016MNRAS.456.2505L}) also determined an orbital period of $\sim$2.11~d.
    
    A detailed study of the DH~Cep system was conducted by \citet{1997ApJ...483..439P}. These authors employed a Doppler tomography method and the UV spectra obtained with the International Ultraviolet Explorer (IUE) and identified the spectral type of the primary and secondary as O6V(O5 -- O6.5) and O7V(O6.5 -- O7.5), with the masses of 39 -- 50\,$M_{\odot}$ and 35 -- 45\,$M_{\odot}$, respectively. \citet{1997ApJ...483..439P} derived an upper limit on the inclination of $i < 43\degr$ based on the combined absolute visual magnitude from the position on the Hertzsprung–Russell (HR) diagram. They also derived a lower limit of $i > 39\degr$  and showed that below this value, the implied spectroscopic masses would become much larger than the evolutionary ones. 

    
    Moreover, being a hot and massive binary system, DH~Cep is of particular interest to the field of  high-energy astrophysics. The observations of this binary made in X-rays by \citet{2010NewA...15..755B} determined $\log(L_{\rm X}/L_{\rm bol}) = -6.7$ in the 0.3 -- 7.5~keV energy band. This latter study also suggested the presence of colliding winds and  the presence of both a cool ($<1$~keV) and a hot ($>1.89$~keV) component, which are possibly associated with the instabilities in the radiation-driven wind shocks.
    
    As DH~Cep is a luminous object across different parts of the spectrum and is situated in a well-known open cluster, it is a relatively well-studied object of the northern sky. However, to date, there has only been one broadband polarimetric study of this binary system that probed a possible phase dependence of DH~Cep linear polarization. The results of this study were reported by \citet{1991ApJ...366..308C}, who suggested the existence of a substantial aspherical envelope of ionized material around at least one of the binary stars, but failed to detect a phase dependence of the observed polarization variability. The main obstacle preventing a firm detection of phase-dependent variations is the relatively low accuracy of polarization measurements ($\sim$$0.09\%$). Recently, \citet{2019BSRSL..88..287A} conducted broadband polarimetric observations of DH~Cep for only six nights, reporting that the average intrinsic linear polarization arising from DH~Cep is less than 1\%. These authors suggest that the degree and angle of the polarization appear to be dependent on the orbital phase. 
    

    \begin{figure*}[htp!]
    \centering
    \includegraphics[width=1\textwidth]{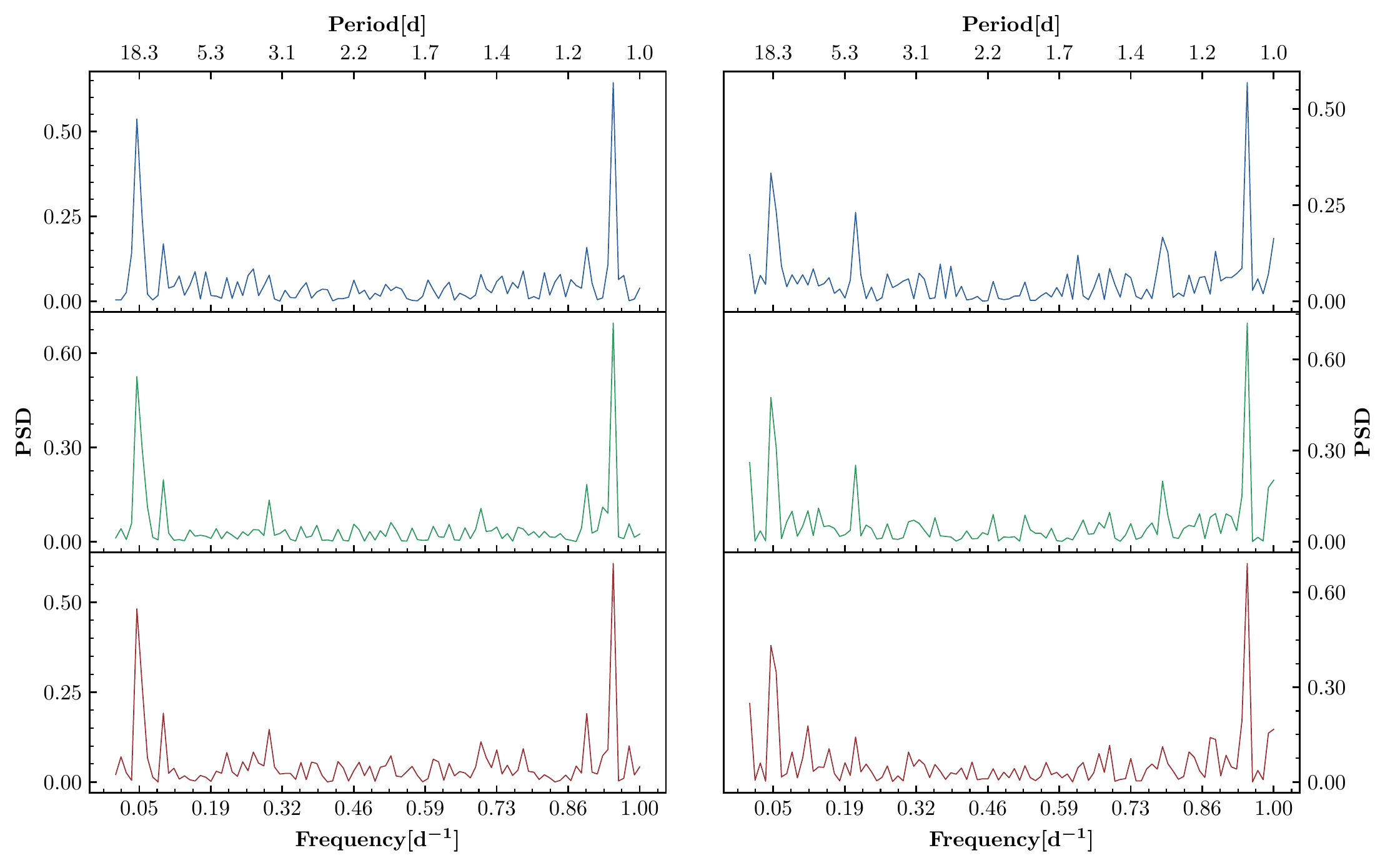}
    \caption{Lomb-Scargle periodograms for Stokes $q$ (left column) and $u$ (right column) of DH~Cep in $B$, $V$, and $R$ passbands (top, middle and bottom panels respectively).} 
    \label{fig:ls}
    \end{figure*}

    In this paper, we present the results of our extensive high-precision polarimetric observational campaign, which was undertaken to probe phase-dependent variability in DH~Cep. Moreover, in order to estimate the polarization in the vicinity of DH~Cep, we measured  the polarization from a sample of field stars identified as members of NGC~7380 and in close proximity to DH~Cep. Our star selection is based on the findings of \citet{1971A&A....13...30M}, who conducted an extensive photometric study of 55 stars in the NGC~7380 cluster. We have chosen 14 stars from the list provided by this latter author, with distances similar to that of DH Cep. This not only gives us an estimate of interstellar polarization in the direction of DH~Cep, but also helps us to understand the possible source of the observed polarization of DH Cep itself. Moreover, we explore the variation of the degree of polarization with the reddening of these stars in the cluster.
        
    \section{Polarimetric observations}\label{sec:observations}
    DH~Cep was observed over 42 nights, from 29 September until 29 December, 2017, with the DiPol-2 polarimeter \citep{2014SPIE.9147E..8IP} installed on the remotely controlled T60 telescope at Haleakal\={a} Observatory, Hawaii. The observational log is given in Table \ref{table:log}.   
     
    DiPol-2 uses two dichroic beam-splitters to split the incident light beam into the $B$, $V$, and $R$ passbands, which are simultaneously recorded by three CCD cameras. The polarimeter uses a superachromatic half-wave plate (HWP) as the polarization modulator and a plane-parallel calcite plate as the polarization analyzer. When observing bright stars such as DH~Cep, DiPol-2 employs an intentional defocusing technique, which allows the collection of up to $10^{7}$ photo-electrons per exposure, avoiding pixel saturation. For DH~Cep, we acquired around 128 -- 192 images every night with an exposure time of 5 -- 7~s for each image, taken at different orientations of the HWP. This translates into 32 -- 48 measurements of the Stokes $q$ and $u$ per night. We acquired a series of dark and bias images once per night. Skyflat images were obtained once per week, on average, in twilight hours, either at the beginning of the observing night or at dawn.

    Our data reduction process involves a standard calibration algorithm with subtraction of bias and dark frames. Our polarimetry algorithm with DiPol-2 automatically eliminates flat-field effects over the areas exposed by the target images. We found that calibration with additional flat-field images does not considerably improve the results. 
    After calibration, normalized Stokes $q$ and $u$ parameters from the flux intensity ratios of the orthogonally polarized stellar images $Q_{\rm i} = I_{\rm e}(i)/I_{\rm o}(i)$ are obtained for each orientation of the polarization modulator at $i$ = $0.0^{\circ}$, $22.5^{\circ}$, $45.0^{\circ}$, $67.5^{\circ}$, and so on. The details of our data reduction procedure can be found in \citet{2020A&A...635A..46P} and \citet{2023A&A...670A.176A}.
    
    In order to determine the instrumental polarization, we observed 20 -- 25 nonpolarized standard stars. The instrumental polarization in the $B$, $V$, and $R$ bands was in the range of $0.004\% - 0.006\%$ and the accuracy of the determination of instrumental polarization is $0.0002\% - 0.0003\%$. To calibrate the polarization angle zero-point, we observed highly polarized standard stars HD 204827 and HD 25443 (see Table \ref{table:hp} for their polarization degrees and angles). The resulting accuracy of the nightly average values of polarization measurements of DH~Cep is at the level of $0.001\% - 0.004\%$ in the $B$ and $R$ bands and $0.002\% - 0.006\%$ in the $V$ band.

    \section{Data analysis}\label{sec:analysis}

    \subsection{Period search}\label{sec:prdsch}
    To search for periodic signals in our polarimetric data, we employed a Lomb-Scargle algorithm \citep{1976Ap&SS..39..447L, 1982ApJ...263..835S}. The benefit of using a Lomb-Scargle algorithm is that it uses the least squares method to fit a sinusoidal function to unevenly sampled data, which is often the case with astronomical data. To execute the Lomb-Scargle algorithm in \textsc{Python}, we used the package from \texttt{astropy.timeseries}\footnote{\url{https://docs.astropy.org/en/stable/timeseries/ lombscargle.html}} \citep{2018zndo...1211397P}. 

    The variations of the Stokes parameters of polarization arising from the scattered light in the binary system typically show two maxima and minima per orbital period separated by 0.5 in the orbital phase \citep{1978A&A....68..415B}. Therefore, the Lomb-Scargle algorithm applied to polarization data is expected to detect half of the orbital period instead of the actual orbital period. We plotted the Lomb-Scargle periodograms for both Stokes $q$ and $u$ in the $B$, $V$, and $R$ passbands in Figure \ref{fig:ls}. There are two clear, prominent peaks in each of these periodograms: at the frequencies of $\sim$0.94~d$^{-1}$ and $\sim$0.06~d$^{-1}$. 

    The first peak clearly indicates (half) the orbital period of DH~Cep and the second peak is merely an alias of the first peak. An alias peak can appear when the frequency of the orbital period is not less than half of the sampling frequency, that is, the Nyquist frequency ($f_{\rm ny}$). In such a situation, two waves are produced that differ by 1/$f_{\rm ny}$. The (half) orbital period for DH~Cep, of namely $\sim$1.055~d, is at a frequency of $\sim$0.94~d$^{-1}$, which is more than half of the sampling frequency of $\sim$0.47~d$^{-1}$, as we used data from 42 nights distributed over a period of about 90~d. Therefore, alias peaks can be expected at around 1/(1 - 1/1.055) = 19.18~d or at frequency close to $\sim$0.05~d$^{-1}$. As one can see, the alias peaks are quite close to that frequency in our periodograms (cf. \citet{1949IEEEP..37...10S}; \citet{2018ApJS..236...16V}). It is not uncommon to observe such alias peaks in Lomb-Scargle periodograms of astronomical data (e.g., \citet{2018MNRAS.478.4710K}; \citet{2023A&A...670A.176A}. 

    Furthermore, we calculated the false-alarm probability (FAP) for the (half) orbital period of DH~Cep and its alias peaks (see Table \ref{table:fap}) using a bootstrap method \citep{2012ada..confE..16S}. The FAP values for both orbital peaks and alias peaks are very low, because the algorithm cannot distinguish between them. However, we conclude that the only period derived from our new polarization data corresponds to the known orbital period of 2.11~d. We do not detect the presence of any other (real) periodic signal in polarimetric data of DH~Cep.

    \begin{table}
    \caption{FAP for DH~Cep periodograms.} 
    \label{table:fap}
    \centering
    \begin{tabular}[c]{l c c c} 
    \hline\hline 
    & Filter & Period[d] & FAP \\ \hline
    
    Stokes $q$: & $B$ & 1.055 & $8.0 \times 10^{-8}$ \\  
    && 18.53 & $1.8 \times 10^{-7}$ \\  
    & $V$ & 1.055 & $7.1 \times 10^{-11}$ \\  
    && 18.36 & $3.4 \times 10^{-10}$ \\  
    & $R$ & 1.055 & $5.4 \times 10^{-8}$ \\
    && 18.41 & $1.0 \times 10^{-7}$ \\  

    Stokes $u$: & $B$ & 1.055 & $9.1 \times 10^{-10}$ \\  
    && 18.35 & $9.0 \times 10^{-9}$ \\  
    & $V$ & 1.055 & $4.7 \times 10^{-13}$ \\  
    && 18.46 & $3.4 \times 10^{-13}$ \\  
    & $R$ & 1.055 & $1.7 \times 10^{-11}$ \\
    && 18.41 & $5.6 \times 10^{-11}$ \\  

    \hline
    \end{tabular}
    \end{table}

    \subsection{Interstellar polarization}\label{sec:isp}
  
    DH~Cep has a parallax of $0.3397 \pm 0.0138$~mas \citep{2021A&A...649A...1G} and is a distant star located on the Galactic plane. Therefore, the observed polarization $P_{\rm obs}$ of DH~Cep may contain a significant interstellar polarization component $P_{\rm is}$ due to a large amount of interstellar dust along the line of sight. In order to properly quantify $P_{\rm is}$, we selected 14 field stars in the vicinity of DH~Cep with parallaxes measured by the Gaia mission. All these stars were previously identified as cluster members \citep{1971A&A....13...30M}. Table \ref{table:is} provides their SIMBAD\footnote{\url{http://simbad.u-strasbg.fr/simbad/sim-fid}} or Gaia DR3 identifiers \citep{2021A&A...649A...1G}---with reference numbers that are the same as those given by \citet{1971A&A....13...30M}---, coordinates, distances, parallaxes, reddening magnitudes, polarization degrees and angles, and the number of polarization measurements. 
    
\begin{figure*}
    \centering
    \subfloat{
    \centering
    \includegraphics[scale=0.94]{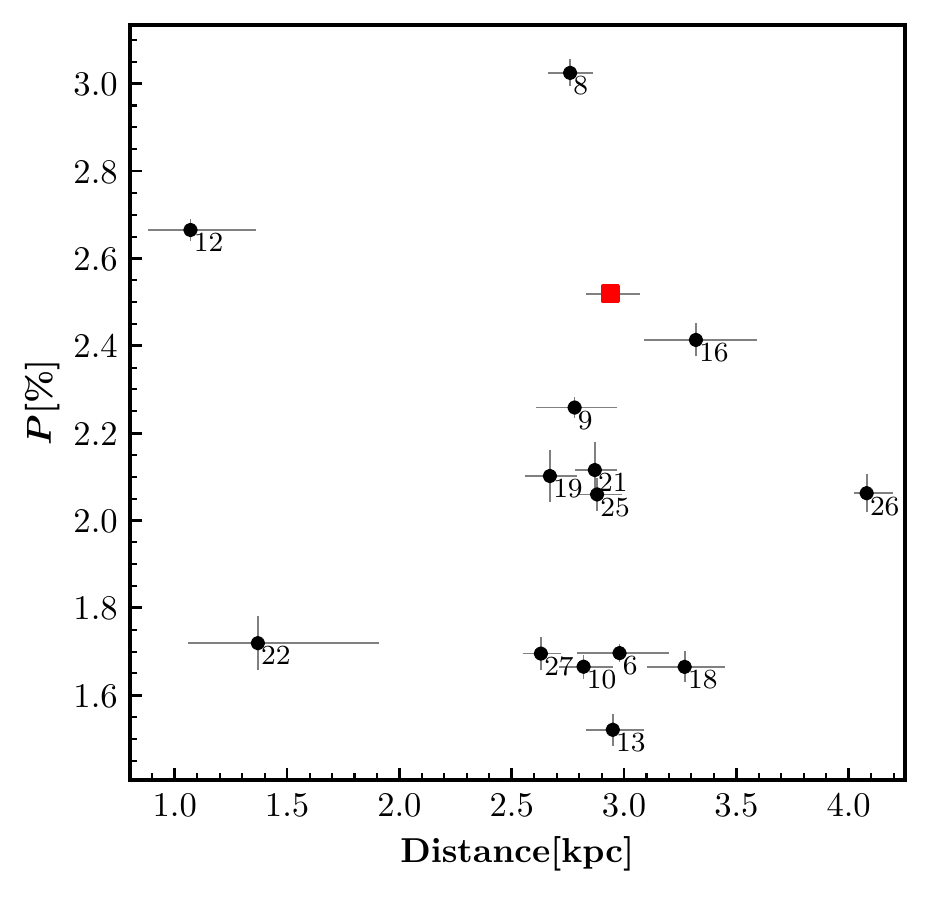}
    }
    \hfill
    \subfloat{\label{}
    \centering

    \includegraphics[scale=0.94]{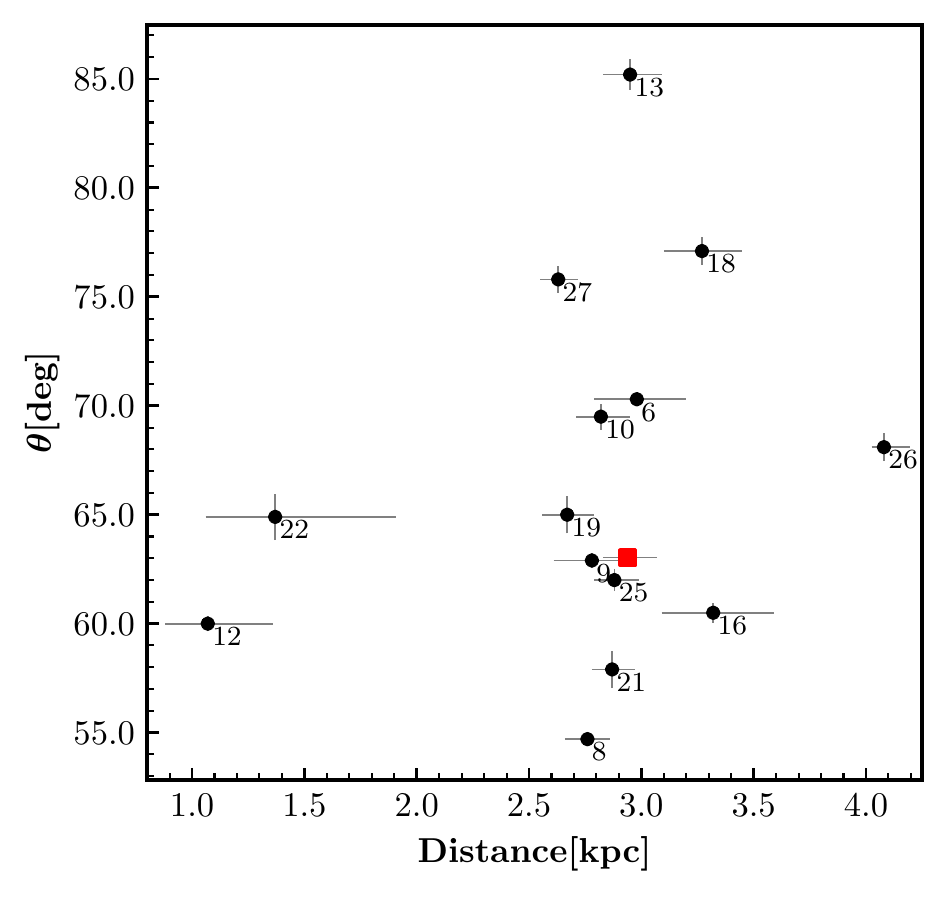}
    }

    \caption{Dependence of the observed degree of polarization $P$ (left panel) and polarization angle $\theta$ (right panel) on distance in the $V$ passband for DH~Cep (red square) and field stars (black circles). Field stars are labeled as in Table \ref{table:is}. Error bars correspond to $\pm\sigma$ errors. The length of the error bar on distance for star 26 is divided by 20 for clarity.}
    \label{fig:refs}
    \end{figure*}

\begin{figure}[!htp]
    \centering
    \includegraphics[scale=0.95]{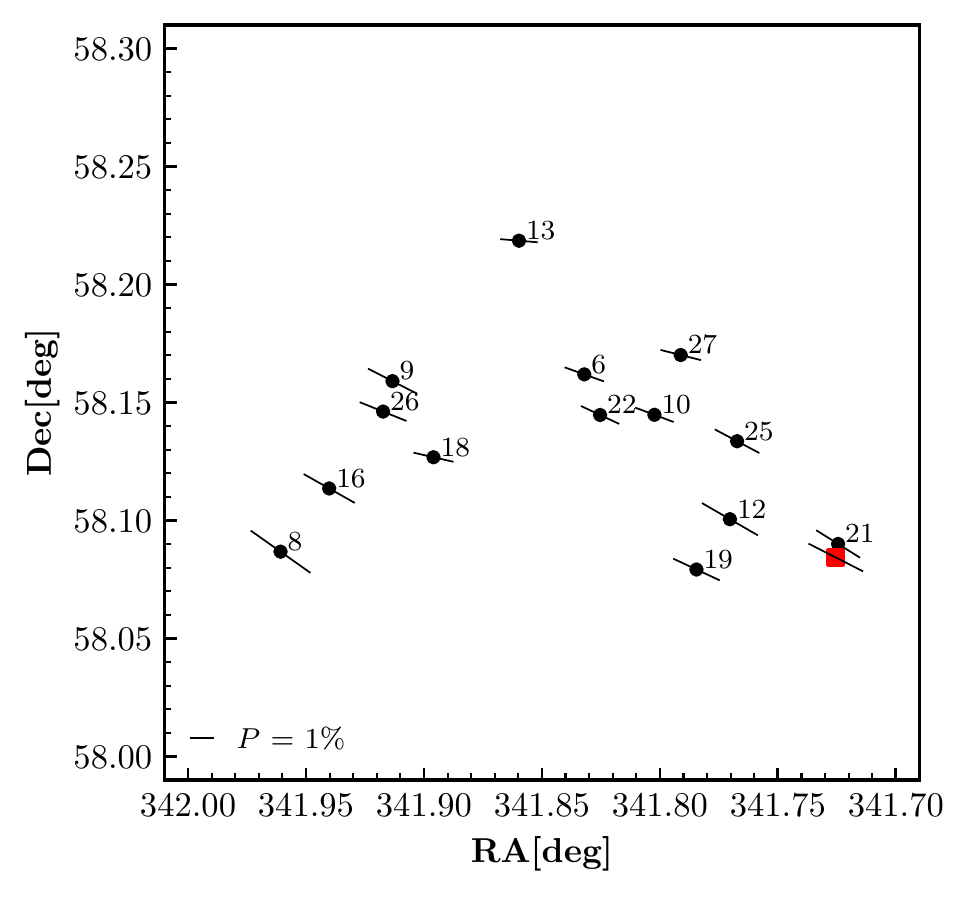}
    \caption{Polarization map of DH~Cep (red square) and field stars (black circles) in $V$ passband. Bar length corresponds to the degree of linear polarization $P$, and the direction corresponds to the polarization angle (measured from the north to the east).} 
    \label{fig:ref}
    \end{figure}

    Figure \ref{fig:refs} shows the dependence of the V-band polarization degree of the field stars and its angle  on distance. We find that two stars from our sample, namely  LS III +57 78 and LS III +57 82, have significantly larger parallaxes and are most likely foreground stars. In Figure \ref{fig:ref}, we depict these field stars together with their values of polarization and polarization angle on the coordinate plane (RA, Dec).

    \begin{table*}
    \centering
    \caption{Identifiers, coordinates, parallaxes, distance, reddening magnitudes, polarization degrees, polarization angles, and the number of polarimetric observations of the field stars.}
    \label{table:is}
    \centering
    \scalebox{1.0}
    {
    \begin{tabular}[c]{ l c c c c c c c c c }
    \hline\hline 
    Identifier  & Coordinates & Distance & Parallax & E(B-V) & Filter & $P$ & $\theta$ & $N_{\rm obs}$ \\
    
    [SIMBAD/Gaia DR3] & [J2000d] & [kpc] & [mas] & [mag] && [\%] & [deg] & \\
    \hline
    
    LS III +57 83 & 341.8320563241, & 2.98 & 0.335 $\pm0.023$ & 0.64 & $B$ & 1.688$\pm$0.017 & $70.5\pm0.3$ & 16 \\
    (Ref 6) & +58.1619232152 &&&& $V$ & 1.696$\pm$0.020 & $70.3\pm0.3$ & 16 \\
    &&&&& $R$ & 1.534$\pm$0.022 & $69.7\pm0.4$ & 16 \\
            
    LS III +57 90 & 341.9607907769, & 2.76 & 0.367$\pm0.013$ & 0.81 & $B$ & 2.772$\pm$0.025 &  $55.8\pm0.3$ & 16 \\
    (Ref 8) & +58.0867661027 &&&& $V$ & 3.025$\pm$0.031 & $54.7\pm0.3$ & 16 \\
    &&&&& $R$ & 2.940$\pm$0.022  & $54.2\pm0.2$ & 16 \\
                
    LS III +57 86 & 341.9133361867, & 2.78 & 0.360$\pm0.023$ & 0.67 & $B$ & 2.259$\pm$0.020 &  $62.0\pm0.3$ & 16 \\
    (Ref 9) &  +58.1590438309 &&&& $V$ & 2.259$\pm$0.024 & $62.9\pm0.3$ & 16 \\
    &&&&& $R$ & 2.102$\pm$0.015 & $63.2\pm0.2$ & 16 \\

    LS III +57 81 & 341.8023077189, & 2.82 & 0.354$\pm0.015$ & 0.54 & $B$ & 1.63$\pm$0.018 &  $68.8\pm0.3$ & 16 \\
    (Ref 10) & +58.1447839479 &&&& $V$ & 1.665$\pm$0.028 & $69.5\pm0.5$ & 16 \\
    &&&&& $R$ & 1.543$\pm$0.015 & $69.4\pm0.3$ & 16 \\
                
    LS III +57 78 & 341.7703133724, & 1.07 & 0.934$\pm0.197$ & 0.50 & $B$ & 2.674$\pm$0.019 &  $58.2\pm0.2$ & 16 \\
    (Ref 12) & +58.1005319096 &&&& $V$ & 2.665$\pm$0.024 & $60.0\pm0.3$ & 16 \\
    &&&&& $R$ & 2.491$\pm$0.027 & $60.4\pm0.3$ & 16 \\
                
    LS III +57 84 & 341.8597287046, & 2.95 & 0.338$\pm0.015$ & 1.01 & $B$ & 1.592$\pm$0.035 &  $82.1\pm0.6$ & 16 \\
    (Ref 13) & 58.2185710310 &&&& $V$ & 1.521$\pm$0.037 & $85.2\pm0.7$  & 16 \\
    &&&&& $R$ & 1.477$\pm$0.028 & $85.1\pm0.5$  & 16 \\
                
    LS III +57 89 & 341.9401580524, & 3.32 & 0.301$\pm0.023$ & 0.61 & $B$ & 2.418$\pm$0.029 &  $61.2\pm0.3$ & 16 \\
    (Ref 16) & +58.1135622187 &&&& $V$ & 2.413$\pm$0.038 & $60.5\pm0.4$  & 16 \\
    &&&&& $R$ & 2.116$\pm$0.023 & $61.1\pm0.3$  & 16 \\

    LS III +57 85 & 341.8959575493, & 3.27 & 0.306$\pm0.016$ & 0.51 & $B$ & 1.699$\pm$0.025 &  $75.9\pm0.4$ & 16 \\
    (Ref 18) & +58.1267941805 &&&& $V$ & 1.665$\pm$0.036 & $77.1\pm0.6$  & 16 \\
    &&&&& $R$ & 1.582$\pm$0.025 & $77.9\pm0.5$  & 16 \\
                
    LS III +57 79 & 341.7845019762, & 2.67 & 0.374$\pm0.016$ & 0.36 & $B$ & 2.191$\pm$0.028 &  $61.9\pm0.4$ & 16 \\
    (Ref 19) & +58.0792187338 &&&& $V$ & 2.102$\pm$0.060 & $65.0\pm0.8$  & 16 \\
    &&&&& $R$ & 2.024$\pm$0.028 & $62.2\pm0.4$  & 16 \\
                
    2007416758659675136 & 341.7244663718, & 2.87 & 0.348$\pm0.012$ & 0.32 & $B$ & 2.046$\pm$0.041 &  $56.2\pm0.6$ & 96 \\
    (Ref 21) & +58.0900466148 &&&& $V$ & 2.115$\pm$0.064 & $57.9\pm0.9$  & 96 \\
    &&&&& $R$ & 2.011$\pm$0.041 & $57.5\pm0.6$  & 96 \\

    LS III +57 82 & 341.8253457807, & 1.37 & 0.731$\pm0.208$ & 0.56 & $B$ & 1.591$\pm$0.045 &  $62.1\pm0.8$ & 15 \\
    (Ref 22) & +58.1447160050 &&&& $V$ & 1.719$\pm$0.061 & $64.9\pm1.0$  & 15 \\
    &&&&& $R$ & 1.623$\pm$0.032 & $65.4\pm0.6$  & 15 \\
                
    2007422840333408256 & 341.7672786883, & 2.88 & 0.347$\pm0.012$ & 0.52 & $B$ & 2.011$\pm$0.036 &  $61.8\pm0.5$ & 16 \\
    (Ref 25) & +58.1335887965 &&&& $V$ & 2.060$\pm$0.037 & $62.0\pm0.5$  & 16 \\
    &&&&& $R$ & 1.882$\pm$0.030 & $64.7\pm0.5$  & 16 \\

    LS III +57 87 & 341.9173203670, & 4.08 & 0.245$\pm0.089$ & 0.29 & $B$ & 2.122$\pm$0.037 &  $67.0\pm0.5$ & 16 \\
    (Ref 26) & +58.1461523123 &&&& $V$ & 2.062$\pm$0.043 & $68.1\pm0.6$  & 16 \\
    &&&&& $R$ & 1.965$\pm$0.030 & $68.5\pm0.4$  & 16 \\
                
    LS III +57 80 & 341.7911729385, & 2.63 & 0.380$\pm0.012$ & 0.40 & $B$ & 1.692$\pm$0.034 & $78.7\pm0.6$ & 16 \\
    (Ref 27) & +58.1701205375 &&&& $V$ & 1.695$\pm$0.037 &  $75.8\pm0.6$  & 16 \\
    &&&&& $R$ & 1.566$\pm$0.030 & $77.1\pm0.5$  & 16 \\
    
    \hline      
    \end{tabular}}
    \end{table*}

    \begin{figure}[!htp]
    \centering
    \includegraphics[scale=0.95]{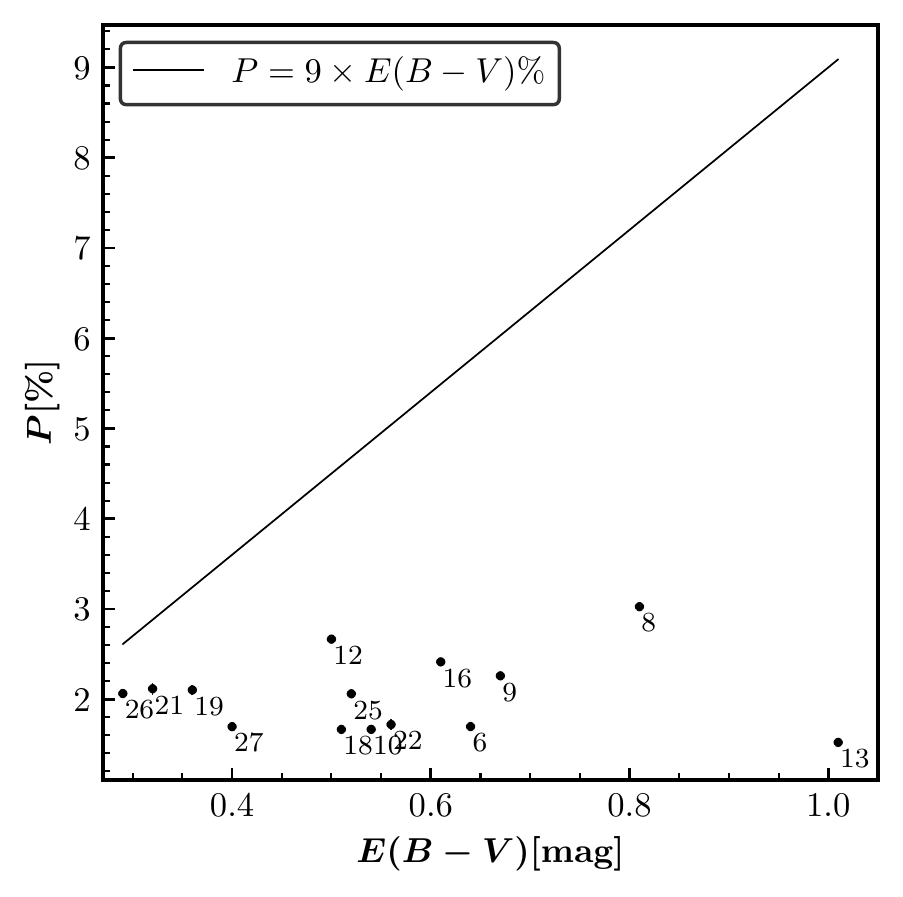}
    \caption{Degree of polarization in the $V$ band plotted versus reddening magnitude for each field star. The straight line in the plot corresponds to $P = 9 \times E(B-V)\%$. For most of the data points on this plot, the $\pm\sigma$ error bars are smaller than their plotting symbols. We note that stars 12 and 22 are foreground stars.} 
    \label{fig:red}
    \end{figure}

    There is a noticeable scatter in the polarization degree (1.5\% -- 3.0\%) in Figure \ref{fig:refs}, even within the narrow distance range of 2.6 -- 3.3~kpc. The direction of polarization of the cluster stars shows only a moderate scatter ($55^{\circ}$ -- $85^{\circ}$), with the average value being around $67^{\circ}$. For all measured stars, the polarization behavior with wavelength shows no peculiarities and is consistent with interstellar polarization. Due to the large scatter in the degree of polarization, using the average value of the polarization computed for all observed cluster stars as an estimate of the interstellar polarization component for DH Cep cannot be justified. However, we believe that the observed polarization values of star 21 can be used as a good first-order estimate. This star is very close to the binary in angular distance ($\sim$$20^{\arcsec}$) and its parallax and distance are the same as those of DH Cep within the errors of determination. Star 21 shows no apparent peculiarities and has no IR excess or emission lines. The direction of its polarization is close to the average direction of polarization in the cluster and its wavelength dependence is consistent with that of interstellar origin. Although we cannot exclude the presence of some intrinsic polarization in star 21, for a nonpeculiar single star to have intrinsic polarization of greater than a level of $\sim$$10^{-4}$ is highly unlikely. Therefore, we adopt the polarization of star 21 as the best available approximation of interstellar polarization for DH Cep itself, keeping in mind the uncertainty due to the inhomogeneity of the interstellar polarization of the cluster member stars.

    Table \ref{table:refs} provides the average observed polarization degree $P_{\rm obs}$ and angle $\theta_{\rm obs}$, the interstellar polarization degree $P_{\rm is}$ and angle $\theta_{\rm is}$ computed for all cluster stars, and intrinsic polarization degree $P_{\rm int}$ and angle $\theta_{\rm int}$ together with their errors for each passband. The latter value is given with two pairs of columns: (a) derived with the average value of the observed polarization for all cluster stars, and (b) derived with the observed polarization of star 21 only. We note that values of $P_{\rm int}$ are very similar for both cases, while the values of $\theta_{\rm int}$  differ by $\sim$$20^{\circ}$. 
   
    The observed average value of polarization in DH~Cep is relatively high ($\sim$2.5\%), but after taking $P_{\rm is}$ into account, the value of the average intrinsic polarization $P_{\rm int}$ is reduced to $\sim$0.6\% with an average of $\theta_{\rm int} \sim 83^{\circ}$. These values are similar to those derived by \citet{1991ApJ...366..308C}, but the estimation of $P_{\rm int} \sim 1\%$ given by \citet{2019BSRSL..88..287A} is somewhat higher than our value. The substantial value of $P_{\rm int}$ supports the conclusion made by \citet{1991ApJ...366..308C} of the presence of a nonspherical circumstellar envelope in this binary system. However, in contrast to that reported by \citet{1991ApJ...366..308C}, we do not see a decrease in $P_{\rm obs}$ in our polarimetric data, which were collected over the period of 90 days.  

    \begin{table*}
    \centering
    \caption{Average observed  $P_{\rm obs}$, $\theta_{\rm obs}$, average interstellar $P_{\rm is}$, $\theta_{\rm is}$, and average intrinsic $P_{\rm int}$, $\theta_{\rm int}$ for DH~Cep.}
    \label{table:refs}
    \scalebox{1.0}
    {
    \begin{tabularx}{\textwidth}[c]{ l c c c c c c c c}
    \hline\hline 
    Filter & $P_{\rm obs}$ & $\theta_{\rm obs}$ & $P_{\rm is}$ & $\theta_{\rm is}$ & $P_{\rm int}\tablefootmark{a}$ & $\theta_{\rm int}\tablefootmark{a}$ & $P_{\rm int}\tablefootmark{b}$ & $\theta_{\rm int}\tablefootmark{b}$ \\
    & [\%] & [deg] & [\%] & [deg] & [\%] & [deg] & [\%] & [deg] \\
    \hline
    $B$ & $2.537\pm0.001$ & $61.5\pm0.1$ & $1.998\pm0.029$ & $66.7\pm0.4$ & $0.665\pm0.029$ & $51.8\pm0.1$ & $0.646\pm0.029$ & $79.2\pm0.1$ \\
                
    $V$ & $2.519\pm0.001$ & $63.0\pm0.1$ & $2.028\pm0.038$ & $67.2\pm0.6$ & $0.583\pm0.038$ & $56.2\pm0.1$ & $0.579\pm0.038$ & $83.4\pm0.1$ \\
                
    $R$ & $2.365\pm0.001$ & $63.5\pm0.1$ & $1.912\pm0.026$ & $67.6\pm0.4$ & $0.547\pm0.026$ & $57.0\pm0.1$ & $0.578\pm0.026$ & $86.7\pm0.1$ \\
    \hline      
    \end{tabularx}}\\
    \tablefoot{
    \tablefoottext{a}{The given values are derived using the average polarization of all observed field stars.}
    \tablefoottext{b}{The given values are derived using the polarization of the field star 21.}} 
    \end{table*}

    Figure \ref{fig:ref} shows these field stars together with their values of polarization and polarization angles on the coordinate plane (RA, Dec). With regard to the interstellar polarization toward NGC~7380 cluster, a couple of attempts to measure this polarization were made in the distant past, providing values of $P_{\rm is} = 1.3\% - 3.0\%$ and $\theta_{\rm is} = 57^{\circ} - 83^{\circ}$ (\citet{1953AJ.....58R..42H}; \citet{1976AJ.....81..970M}). Our estimations are in good agreement with these older values.  Moreover, when comparing polarization data obtained for the same individual stars, our new measurements are in good agreement with those published by \citet{1976AJ.....81..970M}.

    Given the new set of polarization data for NGC~7380, we decided to study the dependence of the interstellar polarization in this cluster on reddening. For this purpose, we plotted the observed degree of linear polarization in the $V$ band for our field stars against their reddening (Figure \ref{fig:red}). For the reddenings, we used the values of $E(B-V)$ given by \citet{1971A&A....13...30M}. As is seen from the plot and in accordance with the conclusions made by \citet{1976AJ.....81..970M}, the polarization of cluster stars most likely arises in the foreground dust layer, and the inter-cluster dust does not polarize. These conclusions are further supported by the fact that the two stars from our sample, LS III +57 78 and LS III +57 82, have significantly larger parallaxes; they are most likely foreground stars, but still show similar polarization degrees and angles to other stars in our sample of field stars. For this reason, there is no apparent increase in $P$ with the increase in $E(B-V)$ (see Figure \ref{fig:red}). We can also see that the polarization of every field star follows the empirical interstellar medium polarization law: $P \le 9 \times E(B-V)\%$ \citep{1975ApJ...196..261S}.

    \subsection{Polarization variability}\label{sec:fcm}

\begin{figure*}
    \centering
    \includegraphics[width=1\textwidth]{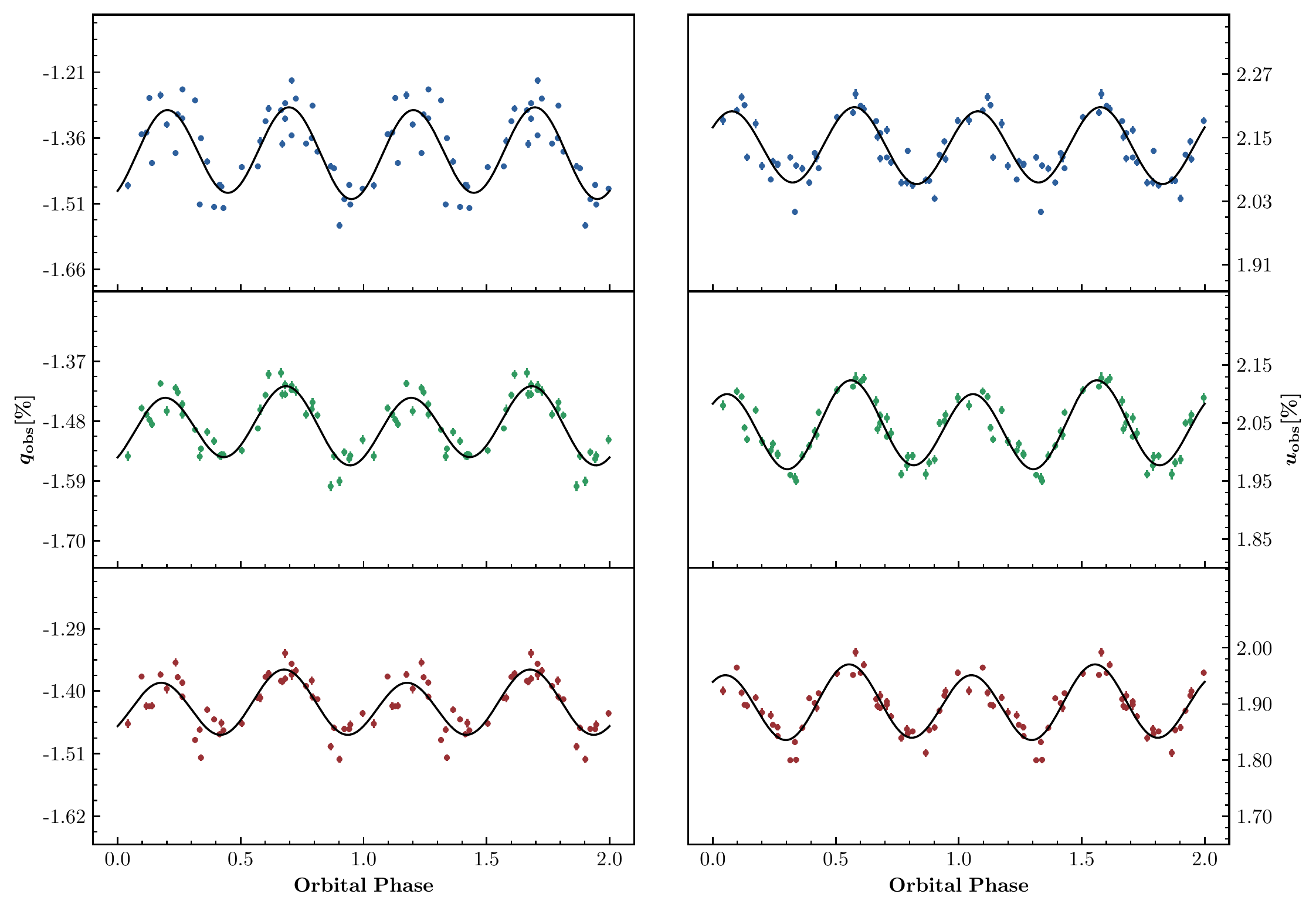}
    \caption{Variability of observed Stokes parameters $q$ and $u$ for DH~Cep in $B$, $V$, and $R$ passbands (top, middle, and bottom panels respectively) phase-folded with the orbital period of 2.11~d. Fourier fit curves (see Sect. \ref{sec:fcm}) are shown with solid lines, and the best-fit Fourier coefficients are given in Table \ref{table:fc}. The error bars correspond to $\pm\sigma$ errors.} 
                \label{fig:fs}
    \end{figure*}

    We used a standard analytical method based on a two-harmonic Fourier fit (commonly known as the Brown McLean Emslie (BME) approach \citep{1978A&A....68..415B}) to fit the phase-folded curves of the Stokes $q$ and $u$ parameters. This method assumes a circular orbit with a co-rotating light-scattering envelope and the fit includes zeroth, first, and second harmonics terms:

    \begin{equation}
    \begin{split}
    q =  q_0 + q_1 \cos \lambda + q_2 \sin \lambda + q_3 \cos 2\lambda + q_4 \sin 2\lambda, \\
    u =  u_0 + u_1 \cos \lambda + q_2 \sin \lambda + q_3 \cos 2\lambda + q_4 \sin 2\lambda,
    \end{split}
    \label{Fourier Series}
    \end{equation}
        
    where $\lambda =  2 \pi \phi$ and $\phi$ is a phase of the orbital period. The polarimetric data of Stokes $q$ and $u$ were phase folded using the orbital period: $P_{\rm orb}~[\rm d] = 2.11095$, and ephemeris: $T_0~[\rm MJD] = 2456525.564$ \citep{2017A&A...607A..82M}, which refers to the time of the conjunction (primary in front). We employed the \texttt{curve$\_$fit} function of the \texttt{scipy.optimize}\footnote{\url{https://docs.scipy.org/doc/scipy/reference/optimize.html}} library in \textsc{Python} to obtain best-fit coefficients together with their errors, as shown in Table \ref{table:fc}. With these coefficients, we performed the curve fitting of the observed polarization data (plots are shown in Figure \ref{fig:fs}).

    As is seen from the fit, the amplitude of the variability is $\sim$0.2\% with noticeable nonperiodic scatter, which is most pronounced in the Stokes $q$ and in the B-band. Revealing such small-amplitude periodic variability with confidence requires precision at the level of $\le 0.03\%$. It appears that the main reason for the nondetection of phase-dependent polarization variability in DH~Cep by \citet{1991ApJ...366..308C} was the relatively low accuracy of polarization measurements. The much higher accuracy that we achieve here allows us to reveal clear periodic variations, showing that nonperiodic scatter of polarization is real; that is, it does not arise from measurement errors. 
    
    One can deduce the orbital inclination $i$ using  the values of the first ($q_{1,2}$, $u_{1,2}$) and second ($q_{3,4}$, $u_{3,4}$) harmonics terms of Fourier series with equations given by \citet{1986ApJ...304..188D} (see Appendix \ref{sec:drissen}). However, for a circular orbit and distribution of the light scattering material symmetric to the orbital plane, first harmonics terms are negligibly small, and therefore only second harmonics terms can be used in practice. This is clearly the case for DH Cep. In addition to the orbital inclination, the orientation of the orbit on the sky (longitude of ascending node  $\Omega$) can be determined (see Appendix \ref{sec:drissen}). 
        
    $A_{\rm q}$ and $A_{\rm u}$ are the two parameters that define the ratio of the amplitudes of the second and first harmonics for Stokes $q$ and $u,$ respectively. These parameters are effective measures of the degree of symmetry and the concentration of scattering material towards the orbital plane in a circular orbit. Formulae for deriving these parameters are also given in Appendix \ref{sec:drissen}.

    Using Eqs. (B1 -- B4), we derived the values of $i$, $\Omega$, $A_{\rm q}$, and $A_{\rm u}$ for the $B$, $V$, and $R$ passbands, which are given in Table \ref{table:orbpar}. We can definitely say that second harmonics terms are dominant for both Stokes $q$ and $u$ parameters in all three passbands. There are small but apparently nonzero first harmonics in the variability of Stokes $q$ and $u$ seen in the V and R bands, but not to the extent that would explicitly confirm a significant asymmetry of light-scattering material, which was suggested by \citet{1991ApJ...366..308C}. 

    It is well known that due to the noise in polarimetric data arising from measurement uncertainties, the derived values of $i$ from the Fourier fit are always biased towards higher values (\citet{1981MNRAS.194..283A}; \citet{1982MNRAS.198...45S}; \citet{1994MNRAS.267....5W}). The amount of bias depends on the value of true inclination; the lower the true value of $i$, the greater the bias toward higher values. A similar bias toward higher values can be induced by the stochastic noise caused by the intrinsic nonperiodic component of polarimetric variations \citep{2000AJ....120..413M}. This bias also affects $1\sigma$ and $2\sigma$ confidence intervals, which become asymmetric, with the lower border extending toward smaller values of $i$. In the case of small $i$ and $\gamma$ (Eq. 2) values, the lower boundary of the confidence interval may extend to $i = 0^{\circ}$. If this happens, polarimetry may only provide an upper limit on the true value of inclination. Moreover, the width of $\Omega$ confidence interval(s) increases rapidly with decreasing  inclination \citep{1994MNRAS.267....5W}.

    \begin{table*}
    \centering
    \caption{Fourier coefficients for Stokes $q$ and $u$.}
    \label{table:fc}
    \centering
    \scalebox{1.0}{
    \begin{tabular}[c]{ l c c c c c c c c c c }
    \hline\hline 
    {\shortstack{\rm Filter }} & {\shortstack{\rm $q_0$}}  & {\shortstack{\rm $q_1$}} 
    & {\shortstack{\rm $q_2$}} & {\shortstack{\rm $q_3$}} & {\shortstack{\rm $q_4$}}
    & {\shortstack{\rm $u_0$}} & {\shortstack{\rm $u_1$}} & {\shortstack{\rm $u_2$}}
    & {\shortstack{\rm $u_3$}} & {\shortstack{\rm $u_4$}} \\                              
    \hline
    $B$ & -1.3924 & -0.0078 & -0.0008 & -0.0801 & 0.0590
    & 2.1337 & -0.0042 & -0.0005 & 0.0399 & 0.0575 \\
    & $\pm$0.0086 & $\pm$0.0129 & $\pm$0.0109 & $\pm$0.0129 & $\pm$0.0113
    & $\pm$0.0046 & $\pm$0.0070 & $\pm$0.0059 & $\pm$0.0070 & $\pm$0.0061 \\
                                
    $V$ & -1.4911 & -0.0110 & -0.0071 & -0.0456 & 0.0442
    & 2.0432 & -0.0098 & -0.0075 & 0.0503 & 0.0471 \\
    & $\pm$0.0037 & $\pm$0.0055 & $\pm$0.0048 & $\pm$0.0056 & $\pm$0.0048 
    & $\pm$0.0032 & $\pm$0.0047 & $\pm$0.0041 & $\pm$0.0048 & $\pm$0.0041 \\                              
                                
    $R$ & -1.4290 & -0.0052 & -0.0105 & -0.0312 & 0.0410
    & 1.8958 & -0.0084 & -0.0051 & 0.0484 & 0.03787 \\
    & $\pm$0.0041 & $\pm$0.0060 & $\pm$0.0053 & $\pm$0.0062 & $\pm$0.0053 
    & $\pm$0.0032 & $\pm$0.0047 & $\pm$0.0041 & $\pm$0.0048 & $\pm$0.0041 \\                              
    \hline      
    \end{tabular}}
    \end{table*}

    Figure \ref{fig:fs} reveals significant scatter around the fitting curves in all passbands. The fit for the $B$ passband ---particularly for the Stokes $q$--- is the most affected. The value of the standard deviation $\sigma$ for the B-band $q$ fit is $\simeq 0.06,$ which is about twice those for the V- and R-band fits. We emphasize that nonperiodic noise is not due to observational errors, which are smaller than $\le0.005\%$ for all wavelength bands. The presence of real stochastic variability in DH Cep polarization is relatively clear. The natural explanation for this phenomenon is highly clumped radiatively driven stellar wind, which is not uncommon in early-type binaries such as DH Cep (a discussion on the polarization mechanism is presented in Section \ref{sec:discussion}). 
    
    As expected, because of the  larger amplitude of the noise, the inclination derived for the B-band will be the most affected; that is, it will be shifted toward higher values by the bias. As DH~Cep is not an eclipsing binary, the true inclination should be rather small. As mentioned in the previous paragraph, a true lower inclination causes greater bias and more asymmetric confidence interval(s) for derived inclination $i,$ and widens the confidence intervals for $\Omega$. In order to quantify this bias, we derived confidence intervals for $i$ and $\Omega$ using the method proposed by \citet{1994MNRAS.267....5W}. This method uses the special merit parameter $\gamma$, which is defined as 
    \begin{equation}
    \gamma = \left(\frac{A}{\sigma_{\rm p}}\right)^2\frac{N}{2},
    \label{gamma}
    \end{equation}
    where $A$ is the fraction of the amplitude of polarization variability, defined as:     
    \begin{equation}
    A = \frac{|q_{\rm max} - q_{\rm min}| + |u_{\rm max} - u_{\rm min}|}{4},
    \label{amplitude}
    \end{equation}
     where $\sigma_{\rm p}$ is a standard deviation that is determined from the scatter of the observed Stokes parameters around the best-fit curves, $N$ is the number of observations, and $q_{\rm max}$, $q_{\rm min}$, $u_{\rm max}$, $u_{\rm min}$ are the maximum and minimum values of the fitted curve of Stokes parameters $q$ and $u$. Our derived values of $\sigma_{\rm p}$ are 0.055, 0.030, and 0.031 and those of $\gamma$ are 65, 158, and 93 for the $B$, $V$, and $R$ bands, respectively.  
        
    To estimate the bias and confidence intervals for $i$ and $\Omega$, we used the plots given in (\citet{1994MNRAS.267....5W}; Figs. 4 and 6(c) therein, where $\gamma = 75.0$, which is closest to our obtained $\gamma$ values). The resulting de-biased values for $i$ are $64^{\circ}$, $46^{\circ}$, and $55^{\circ}$ in the $B$, $V$ and $R$ passbands, respectively. The corresponding lower limits on the 1$\sigma$ confidence interval for $i$ extend to 0$^{\circ}$ for all passbands, whereas the upper limits are 67$^{\circ}$, 57$^{\circ}$, and 61$^{\circ}$. Similarly, the values for $\Omega$ with a 1$\sigma$ confidence interval are $2^{\circ}(182^{\circ}) \pm55^{\circ}$, $105^{\circ}(285^{\circ}) \pm55^{\circ}$, and $88^{\circ}(268^{\circ}) \pm55^{\circ}$ for the $B$, $V,$ and $R$ bands. It is to be noted that there is ambiguity as to the value of $\Omega$, as $\Omega + 180^{\circ}$ is equally possible \citep{1986ApJ...304..188D}.
  
\begin{figure}
    \centering
    \includegraphics[width=0.5\textwidth]{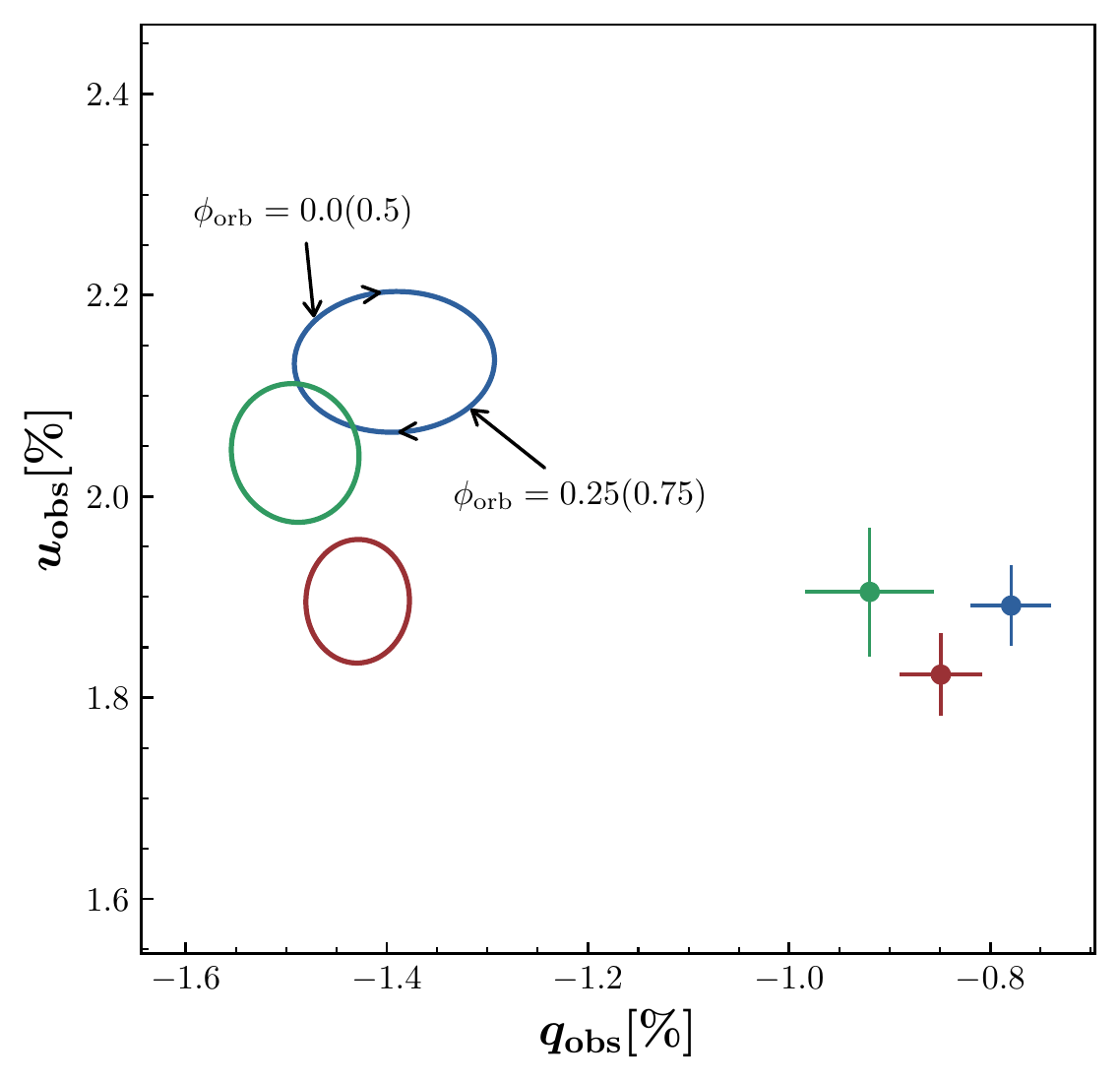}
    \caption{Variability of observed polarization for DH~Cep plotted on the Stokes ($q$, $u$) plane, represented by the ellipses of the second harmonics of the Fourier fit. Blue, green, and red colors represent the $B$, $V$, and $R$ passbands, respectively. The clockwise direction and phases of the orbital period are depicted for the $B$ band ellipse. The angle between the major axis and the $q$-axis gives the orientation $\Omega$. Average interstellar (Ref 21) Stokes $q$ and $u$ parameters are depicted as circles with $\pm\sigma$ error bars.} 
    \label{fig:elps}
    \end{figure}

    Furthermore, we show the ellipses of the second harmonics on the ($q$, $u$) plane in Figure \ref{fig:elps}, which also shows average interstellar polarization for the $B$, $V$, and $R$ bands. The eccentricity of the ellipses is related to the inclination of the orbit $i$, and the orientation of their major semi-axes with respect to the $q$-axis defines the orientation of the orbit $\Omega$. The direction of circumvention corresponds to the direction of the orbital motion in the binary system on the plane of the sky. As one can see, the ellipse for the B-band is oriented almost perpendicular to those for the V and R bands. We believe that this is most likely an effect of the nonperiodic noise on the Fourier fit, which introduces a large uncertainty on the orientation of the second harmonics ellipses on the ($q, u$) plane. Due to the fact that the derived value of inclination is: $90^{\circ} \leq i \leq 180^{\circ}$, the binary system rotation on the plane of the sky is clockwise. The values given in Table \ref{table:orbpar} are subtracted from $180^{\circ}$.
 
    The very first estimation of $i = 62^{\circ}$ was given by \citet{1949AJ.....54..135P} from the analysis of spectroscopic observations. However, such a large inclination value would likely allow eclipses of the two components and this estimate was later rejected. From their photometric observations, \citet{1986IBVS.2932....1L} estimated the inclination to be between $45^{\circ}$ and $55^{\circ}$. Based on their estimation, \citet{1991ApJ...366..308C} adopted a value of $50^{\circ}$. Later on, the photometric observations of \citet{1986IBVS.2932....1L} were modeled by \citet{1994A&A...282...93S} and \citet{1996A&A...314..165H} and both derived $i = 47^{\circ} \pm 1^{\circ}$, which has since been accepted as the inclination value of the orbit of DH~Cep. Our de-biased value of $i = 46^{\circ}$ in the $V$ band, which is our best estimate among three passbands, is very similar to that value. However, our derived value is slightly higher than the upper limit of $i = 43^{\circ}$ provided by \citet{1997ApJ...483..439P}.

    \begin{table}
    \centering
    \caption{Orbital Parameters of DH Cep in $B$, $V$, and $R$ passbands. The parameters $A_{\rm q}$ and $A_{\rm u}$ are defined in Eq. B.4.} 
    \label{table:orbpar}
    \begin{tabular}[c]{ l  c c c c} 
    \hline\hline 
    Filter & $i$\tablefootmark{~a} & $i$\tablefootmark{~b} & $\Omega$ & $A_{\rm q}/A_{\rm u}$ \\ \hline
    $B$ & $66^{\circ}$ & $64^{\circ}$ & $2^{\circ}(182^{\circ})$ & 16.7/12.7 \\  
    $V$ & $50^{\circ}$ & $46^{\circ}$ & $105^{\circ}(285^{\circ})$ & 4.1/5.6 \\  
    $R$ & $57^{\circ}$ & $55^{\circ}$ & $88^{\circ}(268^{\circ})$ & 4.4/6.8 \\  
    \hline
    \end{tabular}\\
    \tablefoot{
    \tablefoottext{a}{The given values are derived using best-fit Fourier coefficients.}
    \tablefoottext{b}{The given values are de-biased.}}
    \end{table} 

    \section{Discussion}\label{sec:discussion}
    The analytical solution of BME does not take into account the finite size of the stars or the eclipses of the scattering material that may take place in close-binary systems. Moreover, this method does not provide detailed information on the distribution of the scattering material that is responsible for the observed polarization variability. Numerical modeling can be employed that takes into account the finite sizes of the components, the distance between them, and the various scattering scenarios that are the most relevant for a given system. As DH Cep is an O-type binary, the most likely cause of the polarization variations is electron scattering in the stellar winds of the hot components. In their study of the stars of the NGC~7380 cluster, \citet{1995ApJ...454..151M} re-classified the components of the DH~Cep binary and added ((f)) to their spectral types, which suggests significant stellar winds. The analysis of the short-wavelength prime (SWP) spectrum of DH~Cep provided by the IUE, which is accessible through the MAST\footnote{\url{https://archive.stsci.edu/index.html}} archive, reveals the unambiguous presence of a strong $\ion{C}{IV}$ $\lambda$1550 wind profile, which indicates that either one or both of the stars exhibit strong wind \citep{1991ApJ...366..308C}. Another possible mechanism is the reflection of light off the facing hemisphere of the components; see for example the case of the O-type binary system LZ Cep \citep{Berdyugin1999}. However, we defer a detailed modeling to our upcoming paper, in which we will perform a numerical modeling of the periodic polarization variability in the early-type systems AO Cas, DH Cep, and HD 165052 in the same way as was already done for the O-type system HD 48099 \citep{2016A&A...591A..92B}.
    
    The polarization variability amplitudes computed from our fit to the data are 0.085, 0.063, and 0.052 \%, in the $B$, $V$, and $R$ passbands, respectively. The scatter due to the nonperiodic component, as determined from the standard deviation, is also larger in the B band. As electron scattering is gray, all amplitudes should be equal irrespective of wavelength. The observed difference in the amplitude may be attributable to the dilution of unpolarized radiation from a larger volume around the binary system, which in turn is due to the increasing amount of free-free continuum emission toward the red. We note that this dilution has the same effect on both periodic and nonperiodic polarization components, explaining the smaller scatter in the $V$ and $R$ passbands.

    Moreover, we deduced the mass-loss rate using the model given by \citet{1988ApJ...330..286S}. This method employs the amplitude ($A_{\rm p}$) of the linear polarimetric variability:

    \begin{equation}
    A_{\rm p} = \frac{(1 + \cos^2 i) 3 \sigma_{\rm t} f_{\rm c} \dot{M}}{(16\pi)^2 m_{\rm p} v_{\infty} a} I, 
    \label{semiamplitude}     
    \end{equation}
    
    where $\dot{M}$ is mass-loss rate, $f_{\rm c}$ is the fraction of the total light coming from the companion star, $v_{\infty}$ is the terminal velocity of the wind, $a$ is semi-major axis, and $I$ is an integral:

    \begin{equation}
    I = \int_{0}^{\infty} \int_{0}^{\pi} \int_{0}^{2\pi} \frac{\sin^{3}\theta \cos 2 \phi 
     d(R/a)d\theta d\phi}{(R^{\prime}/a)^2(1 - R_{*}/R^{\prime})^{\beta}}. 
    \label{integral}     
    \end{equation}

    We note that primed coordinates are measured relative to the primary star, while unprimed coordinates originate at the secondary star. In order to evaluate this integral, one has to choose a specific wind velocity law: $v(R^{\prime}) = v_{\infty}(1-R_{*}/R^{\prime})^{\beta}$,  which is characterized by the parameter $\beta$. We can rewrite Eq. \ref{semiamplitude} as follows:

    \begin{equation}
    \begin{split}   
    & \dot{M} = \frac{(16\pi)^2 m_{\rm p} v_{\infty} a A_{\rm p}}{(1 + \cos^2 i) 3 \sigma_{\rm t} f_{\rm c} I}, \\
    & \textrm{or} \\
    & \dot{M} [M_{\odot} \; {\rm yr^{-1}]} = \frac{2.33 \times 10^{-7} v_{\infty} ({\rm km \; s^{-1}}) a (R_{\odot}) A_{\rm p}}{(1 + \cos^2 i) f_{\rm c} I}.
    \label{mass-loss}     
    \end{split}
    \end{equation}

    Using our results for polarimetric variations in $V$ passband, we obtain $A_{\rm p} = 0.00063$ (we note that the value of $A$ is given earlier in this section as a percentage), and $i = 46^{\circ}$. \citet{1991ApJ...366..308C} gave the terminal wind velocity of the primary as $v_{\rm \infty} = 3000~\rm km \; s^{-1}$. The semi-major axis $a = 6.8~R_{\rm \odot}$ \citep{1997ApJ...483..439P}, while $f_{\rm c} = (1 + 10^{-0.4\Delta M_{\rm v}})^{-1} = 0.475$, for which we used $M_{\rm v} = -4.66$ and $M_{\rm v} = -4.55$ \citep{1996A&A...314..165H} for the primary and the secondary components, respectively. We used the plot given by (\citet{1988ApJ...330..286S}; Fig. 9 therein) in  order to choose the appropriate value of $I = 12.5$, which is obtained for $\beta = 0.8$ as suggested by \citet{1986ApJ...303..239A} and \citet{1991ApJ...366..308C} for O-type stars, $\epsilon = R_{\rm i}^{\prime}/R_{*}= 1.0$ (assuming that the scattering envelope is optically thin), where $R_{\rm i}^{\prime}$ is measured with respect to the center of the primary star, and $a/R_{*} \sim 1$ using  the value of $R_{*} = 8.1~R_{\odot}$ for $i = 47^{\circ}$ given by \citet{1997ApJ...483..439P}. These values yield $ \dot{M} = 1.7 \times 10^{-7} M_{\rm \odot} \; \rm yr^{-1}$ (the value of $A_{\rm p}$ is divided by two as we assume two equal winds in O+O binaries \citep{2008cihw.conf...39S}) for the primary component of DH~Cep, and $\dot{M}$ for the binary system would be about twice this value, which is lower than the mass-loss rate of $\sim$$2 \times 10^{-6} M_{\rm \odot} \; \rm yr^{-1}$ estimated by \citet{1991ApJ...366..308C}.

    \section{Conclusions}\label{sec:conclusions}        
    We conducted the first comprehensive high-precision polarimetric study of the O+O-type binary DH~Cep. This study reveals clear periodically variable linear polarization with an amplitude of $\sim$$0.2\%$. The periodic signal frequency, which corresponds to half of the previously determined orbital period of 2.11~d, was found using the Lomb-Scargle algorithm. In addition to periodic variability, there are significant nonperiodic fluctuations of the intrinsic polarization. The nonperiodic variability component effectively limits the accuracy of estimates of orbital parameters that can be derived from the Fourier fit to polarization data. Furthermore, our study allows us to obtain an estimate of the interstellar polarization component in DH~Cep based on the observed polarization of the closely located field star 21. If this estimate is accurate, most of the DH Cep polarization ($\sim$2.0\%) is of interstellar nature, and this binary system exhibits a $\sim$0.6\% intrinsic polarization component. This supports the presence of a nonspherical circumstellar scattering envelope in DH~Cep system, which was previously suggested by \citet{1991ApJ...366..308C}. However, we do not detect any decrease in the degree of observed polarization on a timescale of three months, which is in contrast to the findings of these latter authors. 


    We find that for the DH Cep system, the second harmonics of the orbital period clearly dominate the observed periodic variability, which suggests a nearly symmetric geometry of the light scattering material with respect to the orbital plane. The orbital parameters derived from our best (smoothest) Fourier fit to the V-band data are: $i = 46^{\circ} + 11^{\circ}/-46^{\circ}$ and $\Omega = 105^{\circ} \pm 55^{\circ}$. Our derived orbital inclination value is in good agreement with the most commonly adopted value of $47^{\circ}\pm 1^{\circ}$. Using this value, together with the polarimetric amplitude in the $V$ passband, we estimate the mass-loss rate in the binary system to be $\sim$$3.4 \times 10^{-7}M_{\rm \odot} \; \rm yr^{-1}$. The direction of motion on the orbit, as seen on the sky, is clockwise.
    
    This study is part of a series of investigations of linear polarization arising from O-type binary systems. We have already published our studies of HD 48099 \citep{2016A&A...591A..92B} and AO Cas \citep{2023A&A...670A.176A} and are presently studying HD~165052. In a future paper of this series, we will present the results of our numerical scattering code modeling for DH~Cep along with AO Cas and HD~165052 systems. This modeling will help to put constraints on the distribution of light-scattering material in these O-type binaries. 

    \begin{acknowledgements}
    This work was supported by the ERC Advanced Grant Hot-Mol ERC-2011-AdG-291659 (www.hotmol.eu). Dipol-2 was built in the cooperation between the University of Turku, Finland, and the Kiepenheuer Institutf\"{u}r Sonnenphysik, Germany, with the support by the Leibniz Association grant SAW-2011-KIS-7. We are grateful to the Institute for Astronomy, University of Hawaii for the observing time allocated for us on the T60 telescope at the Haleakal\={a} Observatory. We are thankful to Vadim Kravtsov for helpful discussions regarding confidence intervals on derived values of orbital parameters. All raw data and calibrations images are available on request from the authors.

    \end{acknowledgements}

    \bibliography{allbib}
    \bibliographystyle{aa}

    \appendix
    \section{Tables}\label{sec:logs}

    \begin{table}[htp!]
    \caption{Log of polarimetric observations for DH~Cep.} 
    \label{table:log}
    \centering
    \begin{tabular}[c]{l  c  c  c }
    \hline\hline 
    Date & MJD & $T_{\rm exp}[\rm s]$ &  $N_{\rm obs}$ \\ \hline
    2017--09--29 & 58025.87 & 800 & 40 \\
    2017--09--30 & 58026.87 & 640 & 32 \\
    2017--10--01 & 58027.89 & 640 & 32 \\
    2017--10--03 & 58029.86 & 640 & 32 \\
    2017--10--05 & 58031.90 & 800 & 40 \\
    2017--10--06 & 58032.87 & 800 & 40 \\
    2017--10--07 & 58033.85 & 800 & 40 \\
    2017--10--08 & 58034.88 & 800 & 40 \\
    2017--10--10 & 58036.80 & 800 & 40 \\
    2017--10--11 & 58037.89 & 800 & 40 \\
    2017--10--18 & 58044.83 & 800 & 40 \\
    2017--10--19 & 58045.82 & 1600 & 40 \\
    2017--10--21 & 58047.82 & 1560 & 39 \\
    2017--10--22 & 58048.83 & 1600 & 40 \\
    2017--10--27 & 58053.78 & 800 & 40 \\
    2017--10--28 & 58054.80 & 1600 & 40 \\
    2017--11--02 & 58059.78 & 800 & 40 \\
    2017--11--03 & 58060.78 & 1120 & 40 \\
    2017--11--05 & 58062.85 & 1792 & 64 \\
    2017--11--06 & 58063.75 & 800 & 40 \\
    2017--11--07 & 58064.74 & 880 & 44 \\
    2017--11--09 & 58066.76 & 1344 & 48 \\
    2017--11--10 & 58067.76 & 1344 & 48 \\
    2017--11--15 & 58072.77 & 1148 & 41 \\
    2017--11--16 & 58073.76 & 1344 & 48 \\
    2017--11--17 & 58074.75 & 1344 & 48 \\
    2017--11--18 & 58075.77 & 1120 & 40 \\
    2017--11--20 & 58077.78 & 1344 & 48 \\
    2017--11--21 & 58078.78 & 1120 & 40 \\
    2017--11--22 & 58079.78 & 1120 & 40 \\
    2017--11--25 & 58082.77 & 1120 & 40 \\
    2017--12--04 & 58091.72 & 1344 & 48 \\
    2017--12--05 & 58092.72 & 1344 & 48 \\
    2017--12--06 & 58093.72 & 1288 & 46 \\
    2017--12--07 & 58094.71 & 1344 & 48 \\
    2017--12--08 & 58095.71 & 1344 & 48 \\
    2017--12--11 & 58098.72 & 1344 & 48 \\
    2017--12--12 & 58099.71 & 1568 & 56 \\
    2017--12--14 & 58101.73 & 1120 & 40 \\
    2017--12--15 & 58102.72 & 1344 & 48 \\
    2017--12--26 & 58113.70 & 1344 & 48 \\
    2017--12--29 & 58116.73 & 1120 & 40 \\
    \hline
    \end{tabular}
    \end{table} 

    \begin{table}
        \caption{Average polarization degrees ($P$), and polarization angles ($\theta$) of highly polarized stars.} 
        \label{table:hp}
        \centering
        \renewcommand{\arraystretch}{1.0} 
        \scalebox{1.0}{
        \begin{tabularx}{0.49\textwidth}[c]{l c c c c} 
                \hline\hline 
                Star & Filter & $P$~[\%] & $\theta$~[deg] & Ref. \\ \hline
                HD~204827  & $B$ & $5.789 \pm 0.011$ & $57.79 \pm 0.02$ & [1] \\ 
                & $V$ & $5.602 \pm 0.019$ & $58.33 \pm 0.02$ & [1] \\  
            & $R$ & $5.079 \pm 0.011$ & $59.21 \pm 0.02$ & [1] \\
                HD 25443 & $B$ & $5.232 \pm 0.092$ & $134.28 \pm 0.51$ & [2] \\
                & $V$ & $5.127 \pm 0.061$ & $134.2 \pm 0.34$ & [2] \\
                & $R$ & $4.734 \pm 0.045$ & $133.65 \pm 0.28$ & [2] \\

                \hline
        \end{tabularx}}
        \tablebib{(1) \citet{2021AJ....161...20P}; (2) \citet{1992AJ....104.1563S}.}
   \end{table} 

   \newpage

    \section{Formulae}\label{sec:drissen}

    The formulae given by \citet{1986ApJ...304..188D} are given in this section. For the orbital inclination ($i$):

    \begin{equation}
    \begin{split}
    \begin{aligned}
    \left(\frac{1 - \cos i}{1 + \cos i}\right)^4 
    & =  \frac{(u_1 + q_2)^2 + (u_2 - q_1)^2}{(u_2 + q_1)^2 + (u_1 - q_2)^2} \\
    & =  \frac{(u_3 + q_4)^2 + (u_4 - q_3)^2}{(u_4 + q_3)^2 + (u_3 - q_4)^2}.           
    \label{Inclination}
    \end{aligned}
    \end{split}
    \end{equation}
    
    The longitude of the ascending node ($\Omega$) can be computed using the following:
        
    \begin{equation}
    \tan \Omega = \frac{A+B}{C+D} = \frac{C-D}{A-B},
    \label{Inclination}
    \end{equation}

    where,
        
    \begin{equation}
    \begin{split}
    A = \frac{u_4 - q_3}{(1 - \cos i)^2}, \:\:\:\:\:\:\:\:\:\: B = \frac{u_4 + q_3}{(1 + \cos i)^2}, \\
    C = \frac{q_4 - u_3}{(1 + \cos i)^2}, \:\:\:\:\:\:\:\:\:\: D = \frac{u_3 + q_4}{(1 - \cos i)^2}.
    \end{split}
    \label{alphabets}
    \end{equation}

    The following formulae can be used to derive $A_{\rm q}$ and $A_{\rm u}$:

    \begin{equation}
    \begin{split}
    A_{\rm q} = \sqrt{\frac{q_3^2 + q_4^2}{q_1^2 + q_2^2}} , \:\:\:\:\: A_{\rm u} = \sqrt{\frac{u_3^2 + u_4^2}{u_1^2 + u_2^2}}. 
    \end{split}
    \label{Ratios}
    \end{equation}

\end{document}